\documentclass[11pt,twoside]{article}

\usepackage{asp2006_soho23}
\usepackage{epsf}
\usepackage{psfig}
\usepackage{lscape}

\begin{document}
\title{The Tale of Two Minima and a Solar Cycle in Between:
An Ongoing Fast Solar Wind Investigation}
\author{M. P. Miralles, S. R. Cranmer,
A. V. Panasyuk, and M. Uzzo}
\affil{Harvard-Smithsonian Center for Astrophysics,
60 Garden Street, Cambridge, MA 02138, USA}

\begin{abstract}
We have measured the physical properties of polar
coronal holes from the minimum activity phase of solar
cycle 23 (1996--1997) to the present minimum
of solar cycle 24 (2007--2009) using the UVCS instrument on SOHO. 
Observations in H~I Lyman alpha
(121.6 nm) and O~VI (103.2, 103.7 nm) provide
spectroscopic diagnostics of proton and O$^{5+}$ bulk
outflow velocities and velocity distributions as a
function of heliocentric distance above the poles of
the Sun. These observations have allowed us to follow
the changes in the physical properties of the polar
coronal holes during solar cycle 23 and its approach
to the current minimum.
Recent ground- and space-based observations have
reported a variety of phenomena associated
with the current minimum. We present the
comparison of observed oxygen line intensities, line ratios,
and profiles for polar coronal holes at both minima
and during solar cycle 23 and show how this new 
minimum manifests itself in the ultraviolet corona.
The comparison of the physical properties of these 
two minima as seen by UVCS in the extended corona, 
now possible for the first time, may provide crucial
empirical constraints on models of extended coronal
heating and acceleration for the fast solar wind.
\end{abstract}

\section{Introduction} 

Polar coronal holes are magnetically open large-scale spatial structures
that are present for most of the solar cycle, and their role is far
from being well understood.
It is clear that strong connections exist between large coronal
holes and the highest-speed wind streams (Krieger et al.\  1973;
Zirker 1977). 

Recent ground- and space-based observations have reported a
variety of phenomena associated with the current solar minimum.
Solar cycle 23 was a longer than expected activity cycle, 
and the current solar cycle 24 minimum did not produce
a quiescent equatorial streamer belt.
The solar wind during the previous sunspot minimum presented a stable
bimodal structure, composed of low-speed streams around the equator
and high-speed streams over the poles.
However, the solar wind from the current minimum period
differs from this typical configuration. During this minimum,
large low-latitude coronal holes, which are also sources of
fast solar wind, were located at
the equator for an extended period of time, unlike in the
previous minimum (see, e.g., Galvin et al.\  2008;
Miralles 2008; Tokumaru et al.\  2009).

The current polar coronal holes and the fast solar wind have
significantly different properties than at the 1996--1997 minimum.
{\em In situ} observations of the solar wind from
both polar coronal holes show that the fast solar wind is slightly
slower, less dense, cooler, and has a lower momentum flux than during
the 1996--1997 solar minimum
(McComas et al.\  2008; Issautier et al.\  2008).
Magnetic field measurements from the Wilcox Solar Observatory
and the Michelson Doppler Imager (MDI) (Scherrer et al.\  1995) on the
{\em{Solar and Heliospheric Observatory}} ({\em{SOHO}})
show that the polar fields are weaker than in the preceding
minimum (Sun et al.\  2008).
In addition, polar hole perimeter measurements using images from
the Extreme ultraviolet Imaging Telescope (EIT)
(Delaboudini\`{e}re et al.\  1995) on {\em SOHO}
indicate a reduction of the coronal hole area in both poles
of about 15\% between 1996 and 2007 (Kirk et al.\  2009).

The exact manner in which the plasma in coronal holes is heated
and accelerated is still unresolved.
An improvement in our understanding of the 
physical processes responsible for the solar wind have come from the
past decade of observations, analysis, and theoretical work
associated with the {\em SOHO} mission.
For example, the importance of magnetohydrodynamic (MHD) waves
has been emphasized by the observations of the Ultraviolet
Coronagraph Spectrometer (UVCS) (Kohl et al.\  1995, 2006)
on {\em SOHO} that heavy ions are
heated to hundreds of times the temperatures of protons and electrons,
and that the velocity distributions are anisotropic,
indicating Alfv\'{e}n wave dissipation via ion cyclotron resonance
(Kohl et al.\  1997, 1998, 2006; Cranmer et al.\  1999, 2008).
These advances were based on observations of
polar coronal holes at the last solar minimum (1996--1997). 
Coronal holes observed since that time have higher electron
densities, lower kinetic temperatures, slower outflow velocities,
and less divergent magnetic field geometries
(Miralles et al.\  2001a,b, 2002, 2004; Miralles 2006). It appears
that the physical processes controlling the extended
heating and acceleration may depend on the density and on the
magnetic field geometry.

This paper provides an overview of the physical properties of the polar
coronal holes during solar cycle 23 and the current cycle 23/24 minimum 
derived from observations by UVCS.
In particular, spectroscopic diagnostics in polar coronal holes
are discussed.  The resulting
plasma properties of these coronal holes are compared.
For a review of the physics of coronal heating and solar wind
acceleration of the associated high-speed solar wind, see
Cranmer et al.\  (2010, these proceedings).

\section{North and South Polar Coronal Holes in Cycle 23}

\begin{figure}[!t]
\epsscale{0.70}
\plotone{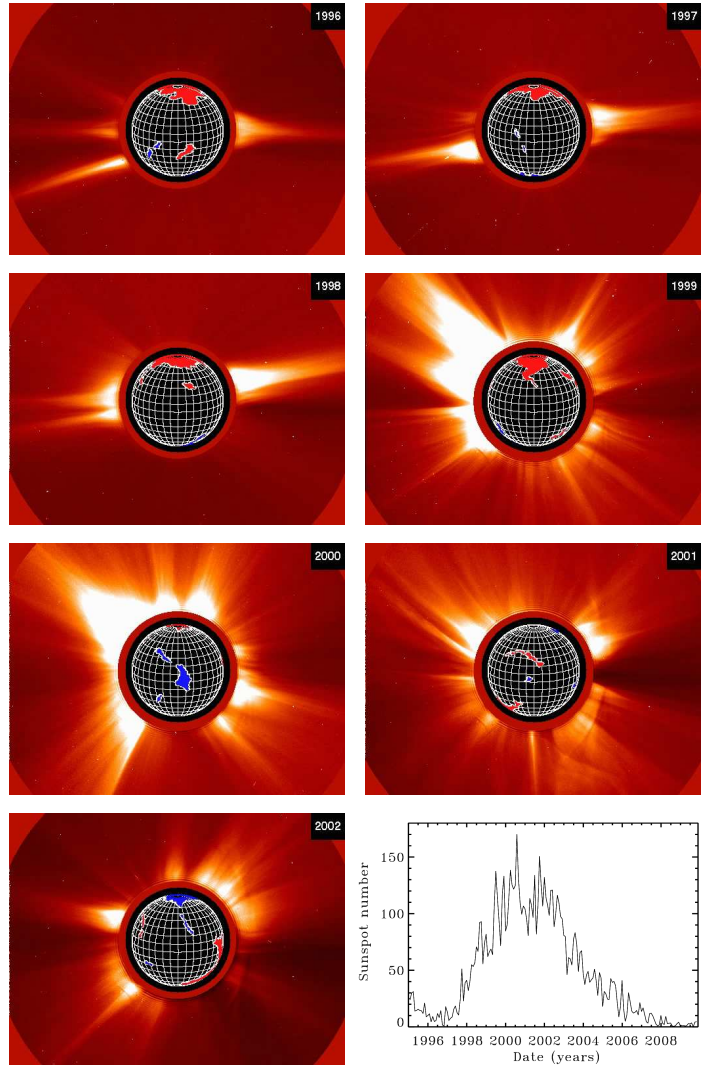}
\caption{Composite representative images of National Solar
Observatory/Kitt Peak Coronal hole boundary He~I 1083 nm (inner)
and {\em{SOHO}}/LASCO C2 (outer) data for December 1996 (CR 1916),
January 1997 (CR 1918), January 1998 (CR 1932),
January 1999 (CR 1945), January 2000 (CR 1958),
January 2001 (CR 1972), and January 2002 (CR 1985), respectively. 
Red (blue) colors on the solar disk represent the positive
(negative) polarity of the field.
Bottom right: Monthly averages of international sunspot number 
from SIDAC, Belgium.  The minimum of solar activity occurred between
1996 and 1997; the maximum of solar activity between 2000 and 2001.}
\end{figure}

Figure 1 illustrates the evolution of polar coronal holes
from 1996 to 2002 with solar cycle sunspot activity.
During the first half of solar cycle 23, the Sun's activity
increased from its lowest level in 1996 to its maximum in 2000,
then decreased again only to rebound in 2001.
This second increase in the Sun's activity level created a double-peaked
activity maximum (Figure 1, bottom-right).
During the solar minimum phase (1996--1997), the
Sun displayed a coronal hole at each of its poles. 
These polar coronal holes were relatively stable structures
that existed for several years (see Figure 1). As the solar
activity increased, the large polar coronal holes shrank and
disappeared near solar maximum 2000, and other smaller coronal
holes emerged at other latitudes (Miralles et al.\  2002, 2004).
These coronal holes of varying size, shape, and polarity 
lasted for several solar rotations.
As the solar cycle continued, the coronal holes completed
their apparent migration to the opposite pole and the Sun's
magnetic polarity reversed.
The times of reappearance of the northern and southern
polar coronal holes differed by over one year in solar cycle 23.
The north polar coronal hole reappeared in February 2001
(Miralles et al.\  2001b, 2002) nearly simultaneously with
the large-scale magnetic polarity reversal of solar cycle 23
(Wang et al.\  2002).
The reformation of the north polar coronal hole with the new
magnetic polarity began with the development of a
high-latitude coronal hole.
It was an elongated structure of $\sim 135^{\circ}$ in longitude
and $\sim 10^{\circ}$--$20^{\circ}$ in latitude
(Miralles et al.\  2001b).
The development of the polar hole and the subsequent expansion into
the north polar region occurred within 8 solar rotations
(Miralles et al.\  2001b). 
At the end of that time, the new polarity coronal hole covered
the pole; it was still large and asymmetric in shape.
The south coronal hole appeared during mid-2002 (Miralles 2003).
By 2006, the polar coronal holes were well developed.

We have used the UVCS instrument to
monitor the evolution of both polar coronal holes from 1996 to present.
The UVCS instrument has been described by Kohl et al.\  (1995, 2006).
The analysis of the coronal hole data set has been described
by Miralles et al.\  (2004).
The in-flight radiometric calibration of UVCS is described by
Gardner et al.\  (1996, 2000, 2002, 2010 these proceedings).
Here we present results for a subset of polar coronal holes, where
foreground and background streamer contributions have been removed
as described by Miralles et al.\  (2001b) and only the broad
coronal emission attributed to the polar coronal hole is shown.
Results for the entire coronal hole data set in solar cycle 23
will be presented in a future publication.

\section{Polar Coronal Hole Properties versus Solar Cycle}

\begin{figure}[!t]
\epsscale{0.70}
\plotone{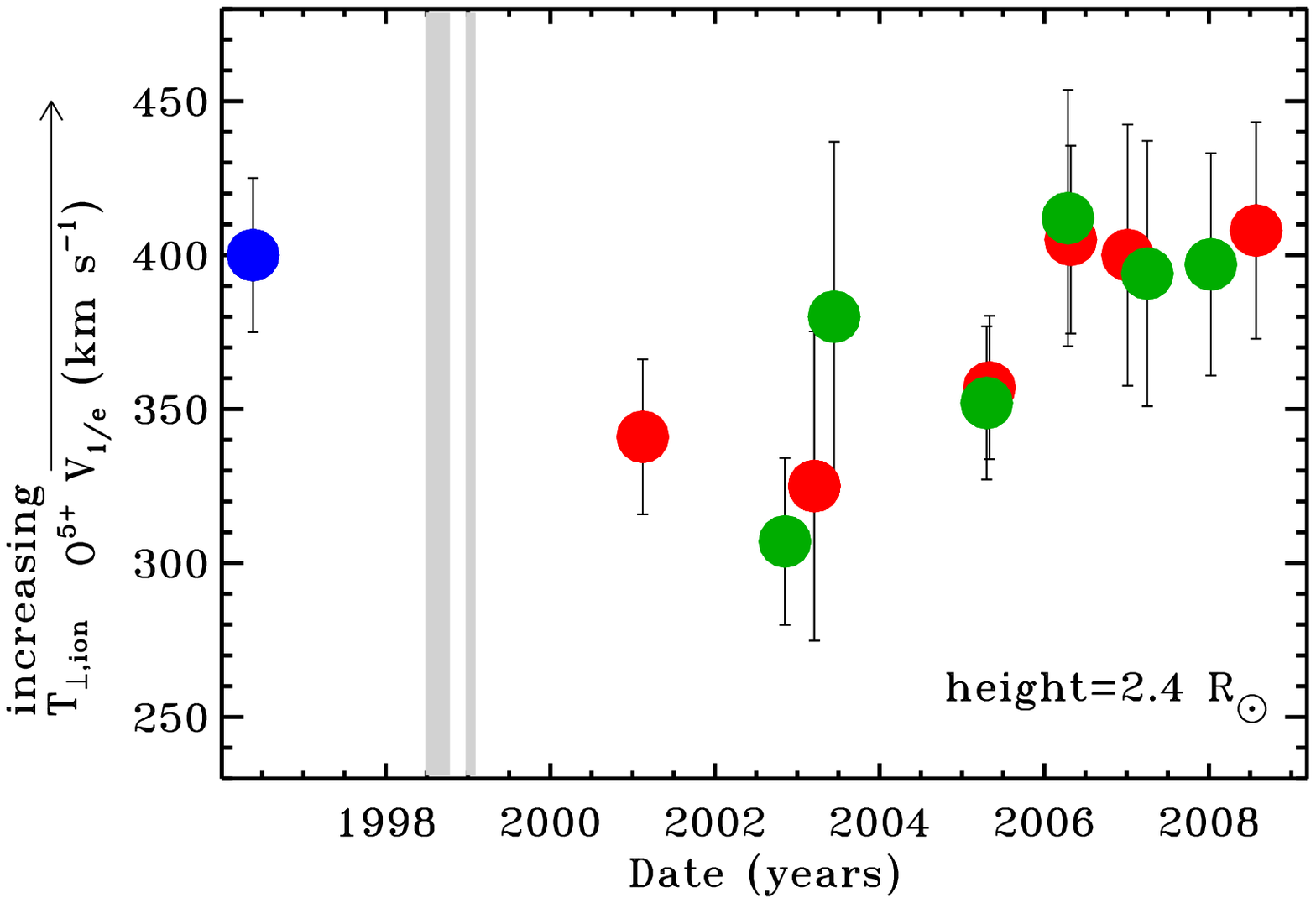}

\epsscale{0.70}
\plotone{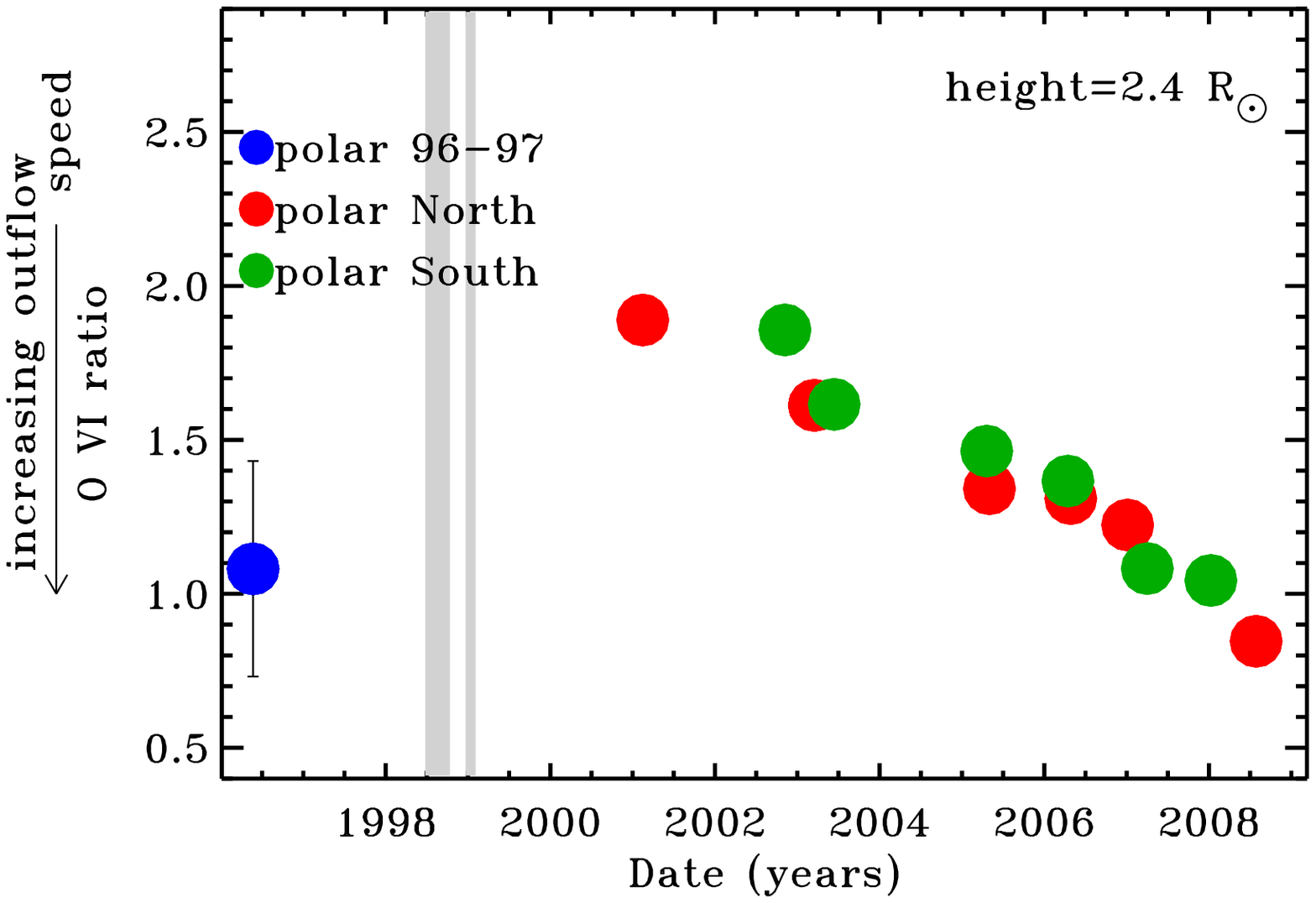}
\caption{O~VI 1032 {\AA} line widths (top) and O~VI line ratio
(bottom) versus time at a height of 2.4 $R_{\odot}$ for polar
coronal holes observed at different phases of solar cycle 23 and
the cycle 23/24 minimum:  1996--1997 solar minimum (blue circles), 
north (red circles), and south (green circles) polar coronal holes.
Gray bands show the interruptions of {\em SOHO} operation.
\label{fig:vstime}}
\end{figure}

Figure~\ref{fig:vstime} shows the variation of O~VI line widths
(top) and the O~VI line ratio (bottom)
with time during solar cycle 23 at a height of 2.4 $R_{\odot}$. 
If we look at the O~VI line width data 
for the polar coronal holes, we can see that at solar minimum,
the line widths were broad and the temperature of the plasma was
the highest.
When the north and south polar coronal holes reformed in early 2001
and mid-2002, respectively, the O~VI line widths were narrower. 
After 2002, the line widths became progressively broader. 
After 2006, the north and south polar coronal holes
started the approach to the extreme plasma temperatures  
measured with UVCS at the last solar minimum (1996--1997).

If we look at the O~VI line ratio for the polar holes at a height of
2.4 $R_{\odot}$, we can see that the lowest
O~VI line ratios (i.e., the highest outflow speeds) were
measured at solar minimum. 
When the north and south polar holes reformed in 2001 and 2002,
respectively, the ratios were at the highest values measured
for polar holes in this cycle implying that the outflow
speeds for those new-polarity polar holes were the lowest.
After 2002, the O~VI line ratios decreased with time
and the outflow speed of the polar holes became larger 
along the cycle.  In early 2007, the north and south polar
coronal holes reached the values of the fast outflow speeds
measured by UVCS at solar minimum in 1996--1997. 

\begin{figure}[!t]
\epsscale{0.70}
\plotone{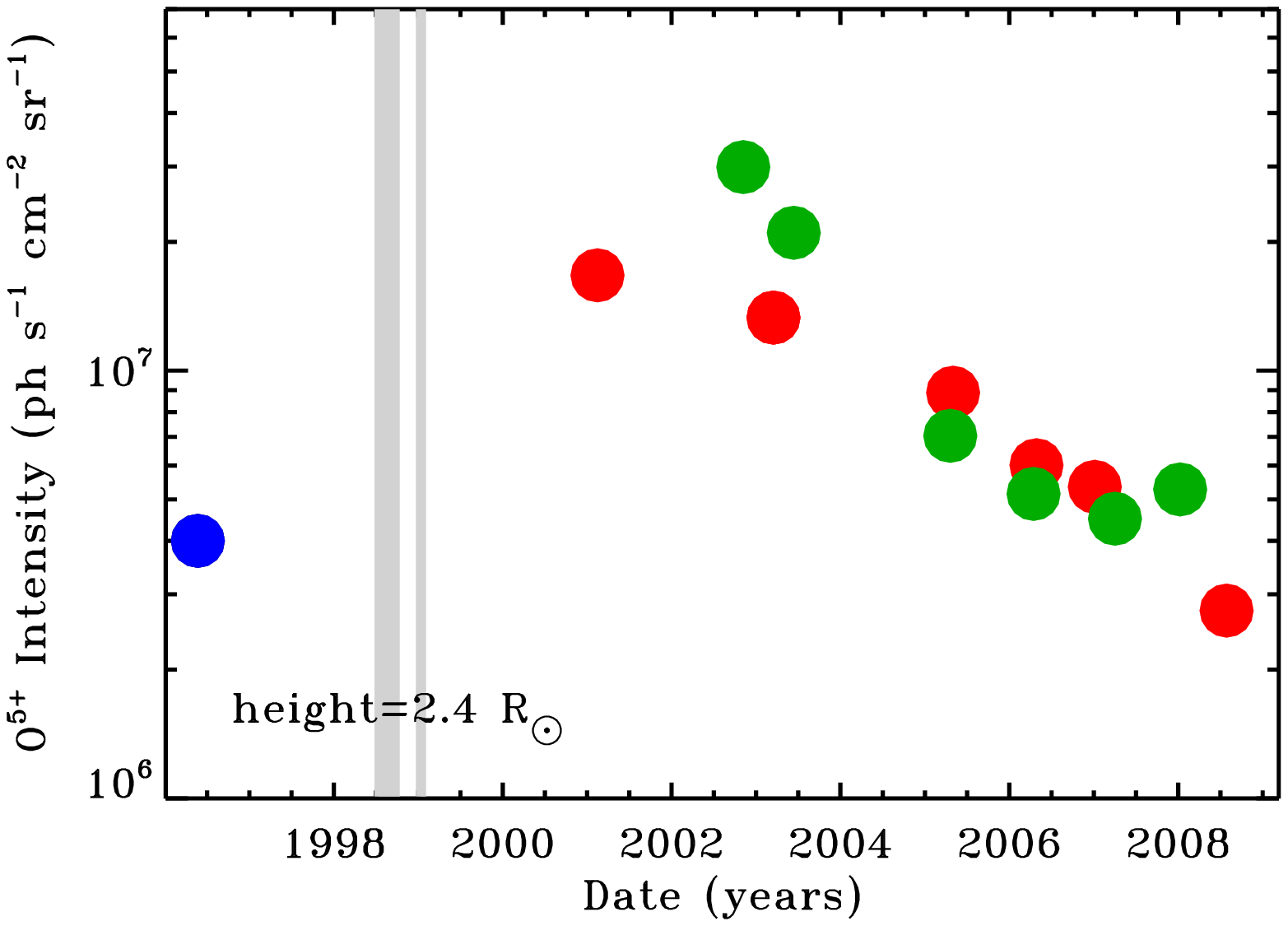}

\epsscale{0.69}
\plotone{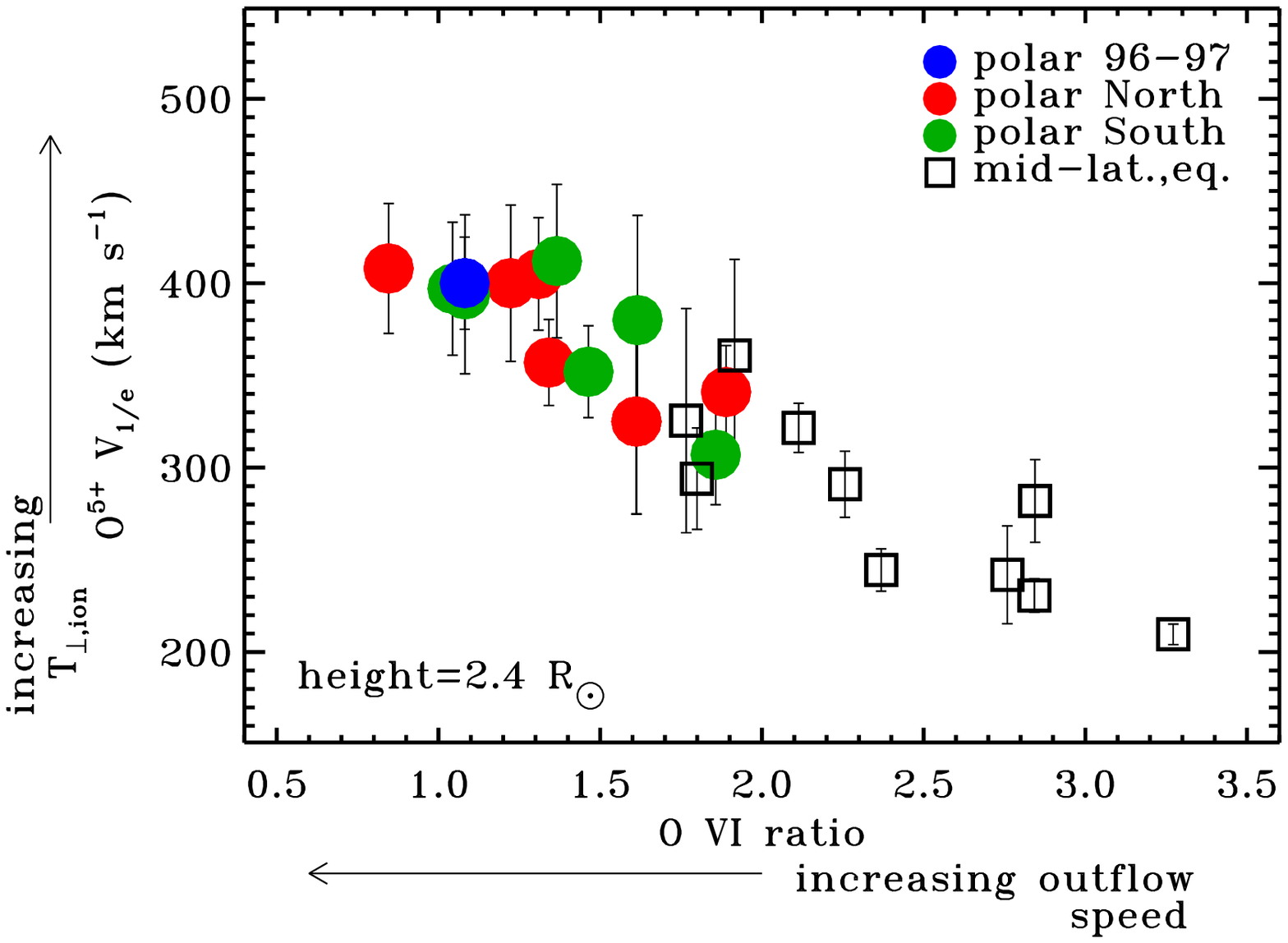}
\caption{O~VI 1032 {\AA} integrated line intensities versus time 
(top) and O~VI 1032 {\AA} line widths versus the O~VI line ratio
(bottom) at a height of 2.4 $R_{\odot}$ for polar coronal holes
observed at different phases of solar cycle 23 and the cycle 23/24
minimum: 1996--1997 solar minimum (blue circles), 
north (red circles), south (green circles) polar coronal holes, and
low-latitude, mid-latitude, and equatorial (squares) coronal holes
are shown.}
\label{fig:linevsratio}
\end{figure}

Figure 3 (top) shows the variation in integrated line intensities
for O~VI 1032 {\AA} in both north and south polar coronal holes
at a height of 2.4 $R_{\odot}$.
After solar maximum, when the north and south polar coronal holes
reappeared, the intensities were higher than at the 1996--1997
solar minimum. After that, the intensities show a downward
trend and reached solar-minimum values by 2007.
In July 2008, there seems to be a decrease in the O~VI 1032 {\AA}
intensities.
This seems to be co-temporal with lower EUV irradiance values observed
toward the end of 2008 (Woods 2010, Didkovsky et al.\  2010,
these proceedings). 
This may be the result of a higher electron temperature or a lower
electron density.
This change in coronal intensities may be consistent with lower 
electron temperatures and densities measured in the heliosphere
(e.g., McComas et al.\  2008).

Figure~\ref{fig:linevsratio} (bottom) shows the O~VI line widths
versus the O~VI line ratio at a height of 2.4 $R_{\odot}$ for large
coronal holes observed at different phases of the solar cycle. 
There seems to be a trend in the physical properties of these
coronal holes.
For large coronal holes, the heavy ions show a strong correlation
between their perpendicular heating and their wind speed  
(Miralles et al.\  2002, 2004).
There are clearly ``zones of avoidance:'' low line intensity
ratios do not occur with the narrowest profiles,
and large line ratios do not occur with the broadest profiles.
In addition, polar coronal holes seem to occupy a different range
in parameter space than low-latitude (equatorial and mid-latitude) 
coronal holes at a height of 2.4 $R_{\odot}$. 
Polar coronal holes have O~VI line ratios between 0.8 and 1.9,
and line widths between 300 and 410 km s$^{-1}$.
Low-latitude holes have O~VI ratios between 1.8 and 3.3,
and line widths between 200 and 360 km s$^{-1}$.
Polar coronal holes show the most extreme plasma
parameters, even at reformation after solar maximum, in comparison to 
other coronal holes at lower latitudes.

\section{Summary}

UVCS/{\em{SOHO}} spectroscopic observations of polar coronal holes
during solar cycle 23 have shown marked variations of ion properties
in the solar wind acceleration region. Polar coronal holes seem to
exhibit different kinetic temperatures and acceleration rates.
These observations show that the north and south polar coronal holes 
in 2007--2009 exhibited similar extreme properties (intensities
and line widths) as those measured by UVCS in the 1996--1997 previous
minimum.
From the analysis of the line widths, we can infer that the
O~VI heating seems as strong as in the previous solar minimum.
A slight decrease in the polar coronal O~VI intensities is
seen that may be consistent with {\em in situ} measurements of
lower electron temperatures and densities in the fast solar wind.
The coronal magnetic field is $\sim$ 40\% lower
(Sheeley 2010, these proceedings), but this magnitude of decrease
in the ion heating may not be present in the corona.

\acknowledgements
This work is supported by NASA grant {NNX\-06\-AG\-95G} to 
the Smithsonian Astrophysical Observatory. 
{\em SOHO} is a joint ESA and NASA mission.


\begin{thebibliography}{}

\bibitem[Cranmer et al.(1999)]{Cranmer99}
Cranmer, S. R., et al. 1999, ApJ, 511, 481

\bibitem[Cranmer et al.(2008)]{Cranmer08}
Cranmer, S. R., Panasyuk, A. V., \& Kohl, J. L. 2008, ApJ, 678, 1480

\bibitem[Delaboudini\`{e}re et al.(1995)]{Delaboudiniere95}
Delaboudini\`{e}re, J. P., et al. 1995, Solar Phys., 162, 291

\bibitem[Galvin et al.(2008)]{Galvin08}
Galvin, A. B., et al. 2008, Eos Trans. AGU, 89 (53),
Spring Mtg. Suppl., SH53A-07

\bibitem[Gardner et al.(1996)]{Gardner96}
Gardner, L. D., et al. 1996, Proc. SPIE, 2831, 2

\bibitem[Gardner, L. D., et al.(2000)]{Gardner00}
Gardner, L. D., Atkins, N., Fineschi, S., Smith, P. L., Kohl, J. L.,
Maccari, L., \& Romoli, M. 2000, Proc. SPIE, 4139, 362

\bibitem[Gardner et al.(2002)]{Gard02}
Gardner, L. D., et al. 2002, in
The Radiometric Calibration of SOHO, ISSI SR-002, eds.
A. Pauluhn, M. C. E. Huber, \& R. von Steiger (Noordwijk,
The Netherlands: ESA), 161

\bibitem[Issautier et al.(2008)]{issau08m}
Issautier, K., Le Chat, G., Meyer-Vernet, N., Moncuquet, M.,
Hoang, S., MacDowall, R.~J., \& McComas, D.~J. 2008,
GRL, 35, L19101

\bibitem[Kirk et al.(2009)]{Kirketal09}
Kirk, M. S., Pesnell, W. D., Young, C. A., \& Hess Webber, S. A.
2009, Solar Phys., 257, 99

\bibitem[Kohl et al.(1995)]{Kohl95m}
Kohl, J. L., et al. 1995, Solar Phys., 162, 313

\bibitem[Kohl et al.(1997)]{Kohletal1997}
Kohl, J. L., et al. 1997, Solar Phys., 175, 613

\bibitem[Kohl et al.(1998)]{Kohl98}
Kohl, J. L., et al. 1998, ApJ, 501, L127

\bibitem[Kohl et al.(2006)]{Kohl06}
Kohl, J. L., Noci, G., Cranmer, S. R., \& Raymond, J. C. 2006,
A\&A Rev., 13, 31

\bibitem[Krieger et al.(1973)]{Krie73}
Krieger, A. S., Timothy, A. F., \& Roelof, E. C. 1973,
Solar Phys., 29, 505

\bibitem[McComas et al.(2008)]{Mc08mm}
McComas, D. J. et al. 2008, GRL, 35, L18103

\bibitem[Miralles(2003)]{Miralles03}
Miralles, M. P. 2003, IUGG 23rd General Assembly,
GAIV.01/02A/A06-002, A.328

\bibitem[Miralles(2005)]{Miralles05}
Miralles, M. P. 2005, IAGA 10th Scientific Assembly,
GAIV.01/IAGA2005-A-01121, 79

\bibitem[Miralles(2008)]{Miralles08}
Miralles, M. P. 2008,
Eos Trans. AGU, 89 (53), Spring Mtg. Suppl., SH53A-04

\bibitem[Miralles et al.(2006)]{Mira06}
Miralles, M. P., Cranmer, S. R., \& Kohl, J. L. 2006, in
SOHO-17: 10 Years of SOHO and Beyond, ed. H. Lacoste, L. Ouwehand
(Noordwijk, The Netherlands: ESA), ESA SP-617, 15.1

\bibitem[Miralles et al.(2001a)]{Mi01a}
Miralles, M. P., Cranmer, S. R., Panasyuk, A. V., Romoli, M., \&
Kohl, J. L. 2001a, ApJ, 549, L257

\bibitem[Miralles et al.(2001b)]{Mi01b}
Miralles, M. P., Cranmer, S. R. \& Kohl, J. L. 2001b, ApJ, 560, L193

\bibitem[Miralles et al.(2002)]{Miralles02}
Miralles, M. P., et al. 2002, in SOHO-11: From Solar Min to Max, ed.
A. Wilson (Noordwijk, The Netherlands: ESA), ESA SP-508, 351

\bibitem[Miralles et al.(2004)]{Miralles04}
Miralles, M. P., Cranmer, S. R., \& Kohl, J. L., 2004,
Adv. Space Res., 33, 696

\bibitem[Scherrer et al.(1995)]{Scherrer95}
Scherrer, P. H., et al. 1995, Solar Phys., 162, 129

\bibitem[Sun et al.(2008)]{Sun08}
Sun, X., Liu, Y., \& Hoeksema, J. T. 2008,
Eos Trans. AGU, 89 (53), Fall Mtg. Suppl., SH51A-1589

\bibitem[Tokumaru et al.(2009)]{Tk09mm}
Tokumaru, M., Kojima, M., Fujiki, K., \& Hayashi, K. 2009, GRL,
36, L09101

\bibitem[Wang et al.(2002)]{Wang02}
Wang, Y.-M., Sheeley, N. R., \& Andrews, M. D. 2002, JGR, 107, 1465

\bibitem[Zirker(1977)]{Zirker77}
Zirker, J. B., ed. 1977,
Coronal Holes and High-Speed Wind Streams (Boulder:
Colorado Assoc.\  Univ.\  Press)

\end{thebibliography}
\end{document}